\documentclass[preprint,showpacs,12pt]{revtex4}
\usepackage{amsmath,amssymb,amsfonts}
\usepackage{graphicx}% Include figure files
\usepackage{hhline}
\usepackage{bm}
\usepackage{amsmath}
\usepackage{epsfig}

\begin{document}         
\begin{flushleft}
CERN-PH-TH/2014- 055\\
\end{flushleft}
\title{Inclusive and pion production neutrino-nucleus cross sections}
%\vspace*{+0.8cm}
\author {M. Martini}
\affiliation{Department of Physics and Astronomy, Ghent University, Proeftuinstraat 86, B-9000 Gent, Belgium}
\author {M. Ericson}
\affiliation{Universit\'e de Lyon, Univ. Lyon 1,
 CNRS/IN2P3, IPN Lyon, F-69622 Villeurbanne Cedex, France}
\affiliation{Physics Department, Theory Unit, CERN, CH-1211 Geneva, Switzerland}

 \begin{abstract}
We analyze the experimental data
 on the  inclusive double differential cross section by neutrinos charged current, measured by T2K,  with the same model which was successful for the MiniBooNE quasielastic cross sections.  As in our previous analysis the multinucleon component is needed in order to reproduce the data. For the total cross section  our evaluation is smaller than the  SciBooNE data above 1 GeV. This  indicates the opening of a new channel not included in our evaluation, presumably the two pion emission channel. We also check that our description holds for the exclusive single pion production channel by confronting our evaluation with the MiniBooNE double differential cross section for a single charged pion and the $Q^2$ distribution. Both are compatible with data. 
\end{abstract}
\maketitle 
%%%%%%%%%%%%%%%%%%%%%%%%%%%%%%%%%%%%%%%%%%%%%%%%%%%%%%%%%%%%%%%%%%%%%%%%%%%%
\section{Introduction}
 Many data on the cross sections of neutrinos or antineutrinos with light nuclei, namely  $^{12}$C, are  now available 
\cite{AguilarArevalo:2010zc,AguilarArevalo:2009ww,AguilarArevalo:2010cx,Nakajima:2010fp,AguilarArevalo:2010bm,AguilarArevalo:2010xt,Abe:2013jth,AguilarArevalo:2013hm}. 
In addition to extensive MiniBooNE results \cite{AguilarArevalo:2010zc,AguilarArevalo:2009ww,AguilarArevalo:2010cx,AguilarArevalo:2010bm,AguilarArevalo:2010xt,AguilarArevalo:2013hm}, 
  the T2K collaboration has issued data on the inclusive double differential cross section \cite{Abe:2013jth}. 
The investigation of this quantity is interesting 
because it brings another test for the validity of a theoretical description. 
Indeed the T2K neutrino beam \cite{Abe:2012av} is different from the MiniBooNE one \cite{AguilarArevalo:2008yp}, it peaks at similar energies, 
$E_\nu \simeq 600$ MeV, as the MiniBooNE one 
but it is definitely narrower and closer to a monochromatic beam. 
Thus the analysis of the T2K 
data offers another test for our description. The differential cross section measured by T2K 
  incorporates  pion production. 
In our work of Ref. \cite{Martini:2009uj} we calculated the quasielastic channel, 
the multinucleon one and 
the single pion (coherent and incoherent) emission. Assuming that these are the only channels  involved in the experiment, the T2K data  are linked to
our total response. 
We also discuss in this work the MiniBooNE data on the double differential partial cross section for single pion production \cite{AguilarArevalo:2010bm}.

 One interesting aspect of these data is the fact that in the angular distributions one bin corresponds 
to small angles for muon emission, $0.95 <\cos \theta <1 $ for MiniBooNE and 
$0.94 <\cos \theta <1$ for T2K. The measurement of the forward cross section offers a chance of access to the elusive isospin spin-longitudinal 
response which is of a particular interest due to its collective aspects with the presence in particular of 
the coherent pion production channel. 
The isospin spin-transverse response, where coherent pion production is essentially absent, quickly dominates when one departs from the forward direction. 
 
In this work we use the same model which has been successful for the MiniBooNE data on the neutrino and antineutrino quasielastic-like cross sections, 
the total or the double differential ones, as shown in Refs. \cite{Martini:2009uj,Martini:2010ex,Martini:2011wp,Martini:2013sha}.  
We summarize here the basic ingredients of this model which is based on the nuclear response functions. In our description 
 the quasielastic response is treated in the random phase approximation (RPA), as discussed by Alberico \textit{et al.} in Ref. \cite{Alberico:1981sz}. 
For the isospin spin-transverse response the particle-hole force is repulsive and 
its  main effect  is a hardening effect and a quenching one due to the mixing of nucleon-hole states with Delta-hole ones, 
the Ericson-Ericson--Lorentz-Lorenz  effect \cite{Ericson:1966fm}. 
The multinucleon contribution is evaluated as in our previous articles \cite{Martini:2009uj,Martini:2010ex,Martini:2011wp,Martini:2013sha}. It is deduced from the microscopic calculation of Alberico \textit{et al.} \cite{Alberico:1983zg} 
on the role of two particle-two hole (2p-2h) contribution in the inclusive $(e,e')$ transverse response. This calculation includes the correlation term,  the  two-body exchange  terms, in particular the one associated with Delta excitation, and the interference between these quantities. 
As for the single pion production we assume, as previously \cite{Martini:2009uj}, that it arises exclusively from the pionic decay of the Delta excitation.   
In the nucleus the Delta width is reduced by medium effects such as the non pionic Delta decay  which leads to 2p-2h or 3p-3h excitations; 
they have been introduced and discussed by Oset and Salcedo in Ref. \cite{Oset:1987re}. We use their parametrization  for the in-medium Delta width. 
The non pionic decay of the Delta in the medium contributes to our n particle-n hole (np-nh) cross section.  
\section{Inclusive cross section} 
\label{sec_inclusive}

\begin{figure}
\begin{center}
  \includegraphics[width=12cm,height=8cm]{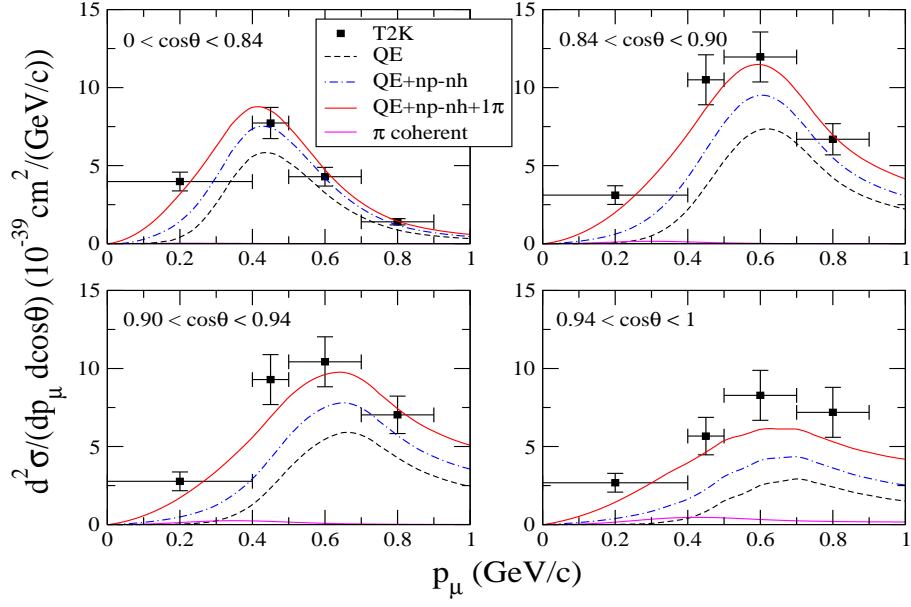}
\caption{(color online). T2K flux-averaged inclusive 
CC double differential cross section on carbon per nucleon
as a function of the muon momentum. The different contributions to this inclusive cross section obtained in our model are shown.   
The experimental T2K points are taken from Ref. \cite{Abe:2013jth}.}
\label{fig_t2k_d2s}
\end{center}
%\vspace*{+0.8cm}
\end{figure}

We first discuss the T2K results \cite{Abe:2013jth}.
The T2K cross section  for charged currents (CC) does not isolate specific channels but sums over all accessible final states. 
For the comparison with these data we assume that the only
channels opened are the quasielastic, the multinucleon emission and the single pion production. 
Notice that here, in contradistinction with the pion emission case, the final state interaction of the emitted pion which depopulates the pion channel, populates
 multinucleon states and therefore does not reduce the inclusive cross section. 
Figure \ref{fig_t2k_d2s} displays our prediction for the double differential cross section as function of the emitted muon momentum, for the various angular bins. 
The experimental points are the T2K measured ones. We show separately the different components of the theoretical cross section: 
first the genuine quasielastic channel, second the total quasielastic-like one including the multinucleon component and the total one including the 
single pion production cross section. The separate contributions are given in order to allow  future comparisons with analysis of the quasielastic channel in T2K which are in progress \cite{Hadley}. The coherent pion production component is also shown in Fig. \ref{fig_t2k_d2s} but in this inclusive cross section its contribution is too small to be singled out. Our evaluation is compatible with the data. As in our previous analysis of the MiniBooNE quasielastic-like cross sections the multinucleon component is needed in order to reproduce the experimental results.

\begin{figure}
\begin{center}
  \includegraphics[width=12cm,height=8cm]{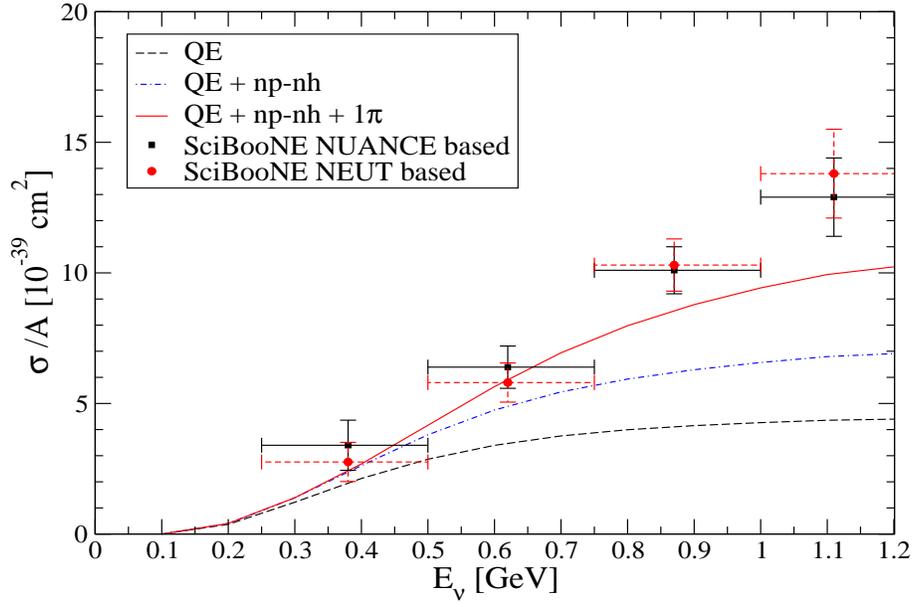}
\caption{(color online). Inclusive 
CC cross section on carbon per nucleon
as a function of the neutrino energy. 
The experimental SciBooNE points are taken from Ref. \cite{Nakajima:2010fp}.}
\label{fig_incl_scib}
\end{center}
%\vspace*{+0.8cm}
\end{figure}

For the smallest angle bin some underevaluation in the theory seems to show up.
 In this respect we can make the following comment. 
 The forward direction, which corresponds to $q \simeq \omega$, is special in one important aspect:
  the spin transverse and the charge (isovector) contributions are kinematically suppressed and only the spin longitudinal one survives \cite{Delorme:1985ps}. 
For small or moderate $q$ values, this last response includes two separated regions of response, 
one at relatively large energy transfers , $\omega>m_\pi$, and one at low energy with the quasielastic component. 
In addition in nuclei the np-nh response fills all the $(\omega,q)$ plane. 
The large energy part contributes to pion emission, coherent or not, and to multinucleon emission. They are included in our predictions. 
The contribution of the low energy part, in the quasielastic region, should in principle be important. 
However in the evaluation of the spin longitudinal contribution there appears a factor $ [\omega - Q^2/(2M)]$, which vanishes 
 identically for the quasielastic kinematics \cite{Marteau:1999kt,Martini:2009uj}. Strictly speaking this cancellation is true for a nucleon initially at rest but 
in practice it remains true also in the Fermi gas. 
Indeed our numerical evaluation of the spin longitudinal quasielastic contribution in neutrino or antineutrino interactions (which is the same in both cases) shows its smallness for all neutrino energies, as illustrated in Fig. 3 of  our previous work \cite{Martini:2010ex}. One can also observe in Fig.\ref{fig_t2k_d2s} 
that the quasielastic contribution is smaller for the smallest angle bin.
In view of these cancellations one is led to consider other contributions beyond the quasielastic kinematics in order to avoid the canceling effect. 
This is for instance the case for the excitation of collective giant resonances. Their energy is  low, $\sim$ 10-30 MeV, which is small compared to the neutrino one, some hundreds of MeV.   The small energy transfer in their excitation implies that the muon energy is nearly the same as the neutrino one, a few hundreds of MeV, the region 
where the excess of the experimental cross section seems to occur. 
At large angles the contribution of the collective states is suppressed by form factor effects 
\cite{Botrugno:2005kn}.  
Several studies have been made on the excitation of low energy collective states in neutrino interactions \cite{Botrugno:2005kn,Kolbe:1995af,Volpe:2000zn,Jachowicz:2002rr,Martini:2007jw,Samana:2010up,Pandey:2013cca} but specific work is needed to assess their role in the present type of data where forward bins offer favorable conditions to display their contribution.

\begin{figure}
\begin{center}
  \includegraphics[width=12cm,height=8cm]{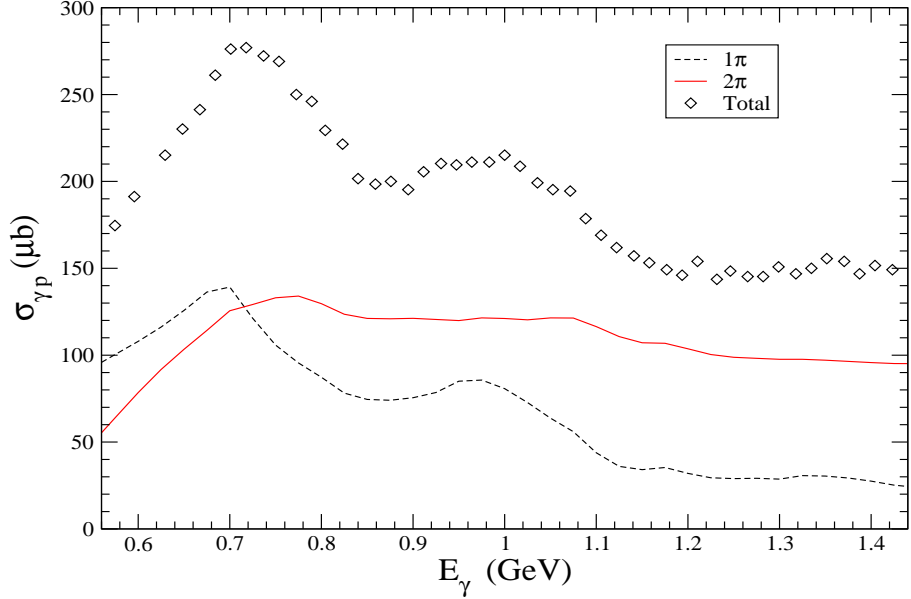}
\caption{(color online). Total photoabsorbtion cross section for free proton as a function of photon energy measured at GRAAL 2008. 
The points are taken from Ref. \cite{graal_pdf}. The one pion production and two pion production contributions are separately plotted.}
\label{fig_photo_abs}
\end{center}
%\vspace*{+0.8cm}
\end{figure}

For the inclusive cross section as a function of the neutrino energy, experimental results have  been previously published by the SciBooNE collaboration 
\cite{Nakajima:2010fp}. We report them in Fig. \ref{fig_incl_scib} together with our theoretical prediction which gives a good fit of the data 
up to  $E_\nu\simeq 1$ GeV  but
 underestimates the cross section above this value, as also reported by Nieves \textit{et al.} \cite{Nieves:2011pp}. The natural interpretation is the existence 
of other channels which open up at high energies and which have not been included in our analysis.  
A likely candidate for the missing channel is the multipion production, in particular the two pion production one, 
as also suggested 
in Refs. \cite{Nieves:2011pp,Buss:2011mx}. 
As an illustration of the likely importance of this channel and although it has no direct connection to the neutrino cross section we report in Fig. \ref{fig_photo_abs} the total photoabsorption cross section by a proton 
as a function of photon energy,
as well as the cross sections for the exclusive channels, one pion production and two pions production ones taken from the GRAAL experimental results \cite{graal_pdf}. Beyond a photon energy of about 
$E_{\gamma }\simeq 0.7$ GeV the two pion production cross section dominates over the single pion one and it represents an important part of the total cross section. 
For incident pions as well, the two pion production cross section which has been studied by Oset and Vicente Vacas \cite{Oset:1986qd} becomes important  
for energies above the Delta resonance and it increases with energy.  For neutrinos as well one can  expect a similar behavior with a dominance of the two pion emission as compared to that of a single pion. The evaluation of the two pion production process by neutrinos has been studied by  
Hernandez \textit{et al.} \cite{Hernandez:2007ej} but only close to the two pion threshold. It should be extended at larger energies. 
A sizable two pion component directly affects the inclusive cross section which sums over final states and 
its omission  is a  likely candidate for  the underevaluation of the total cross section by our theory at large neutrino energies. 
We also remind the possible contribution of deep inelastic scattering, recently examined in connection with the T2K results in Ref. \cite{Lalakulich:2013iaa}.

\section{One pion production cross section} 
\label{sec_pion}
\begin{figure}
\begin{center}
  \includegraphics[width=16cm,height=12cm]{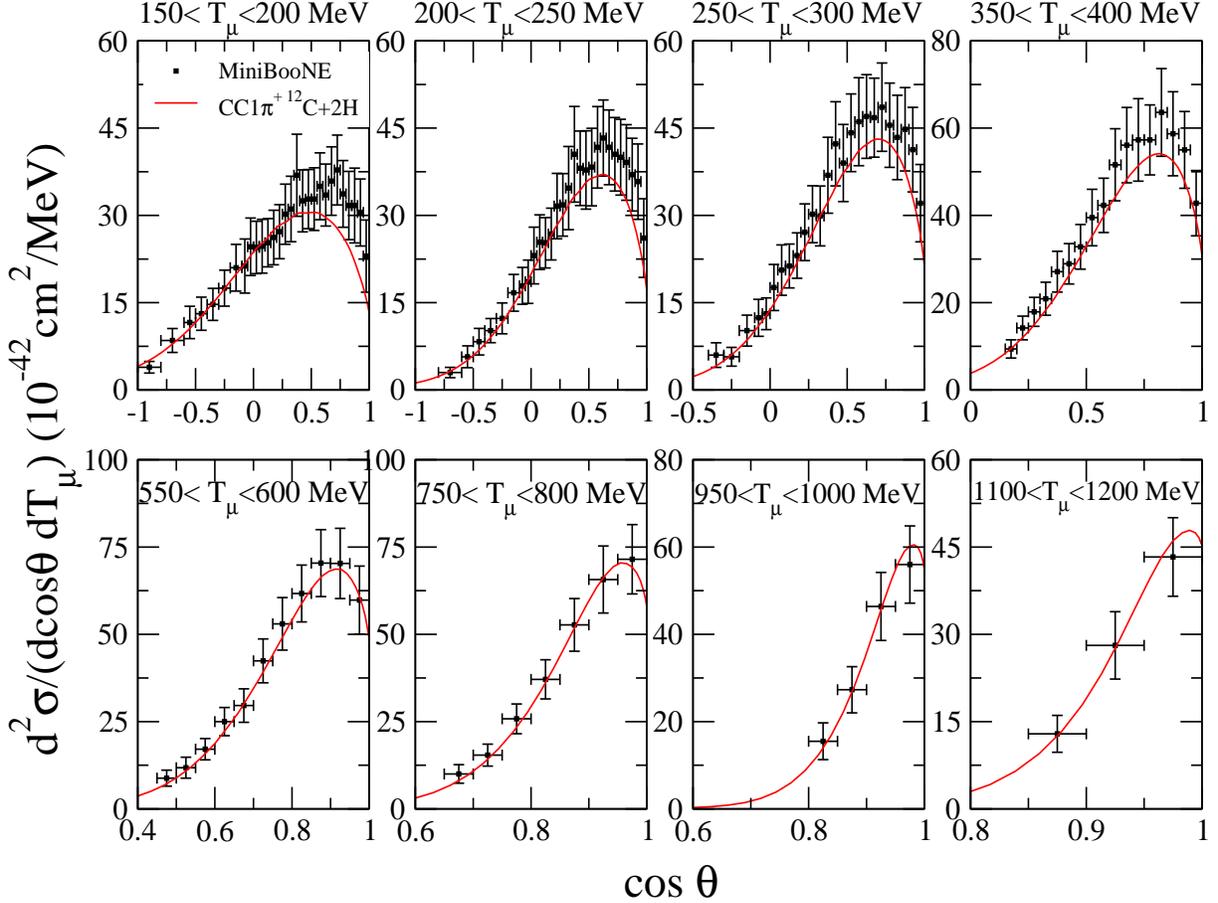}
\caption{(color online). MiniBooNE flux-averaged CC 1$\pi^+$ $\nu_\mu$-CH$_2$ double differential 
cross section for several values of muon kinetic energy
as a function of the muon scattering angle.  
The experimental MiniBooNE points with the shape uncertainty are taken from Ref. \cite{AguilarArevalo:2010bm}.}

\label{fig_minib_d2s_piplus}
\end{center}
%\vspace*{+0.8cm}
\end{figure}

\begin{figure}
\begin{center}
  \includegraphics[width=12cm,height=8cm]{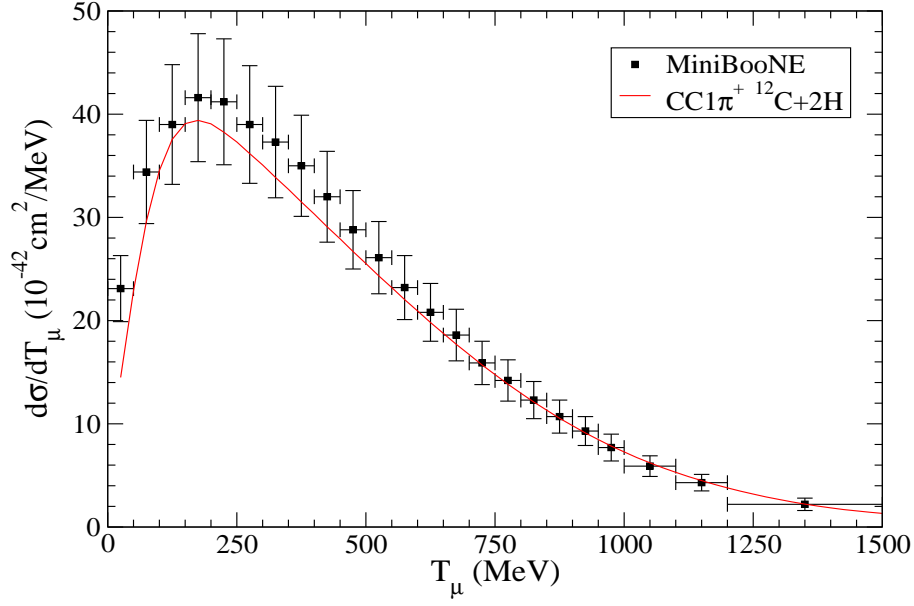}
\caption{(color online). MiniBooNE flux-averaged CC 1$\pi^+$ $\nu_\mu$-CH$_2$ differential cross section 
as a function of the muon kinetic energy. 
The experimental MiniBooNE points are taken from Ref. \cite{AguilarArevalo:2010bm}.}
\label{fig_minib_ds_dt_piplus}
\end{center}
%\vspace*{+0.8cm}
\end{figure}

\begin{figure}
\begin{center}
  \includegraphics[width=12cm,height=8cm]{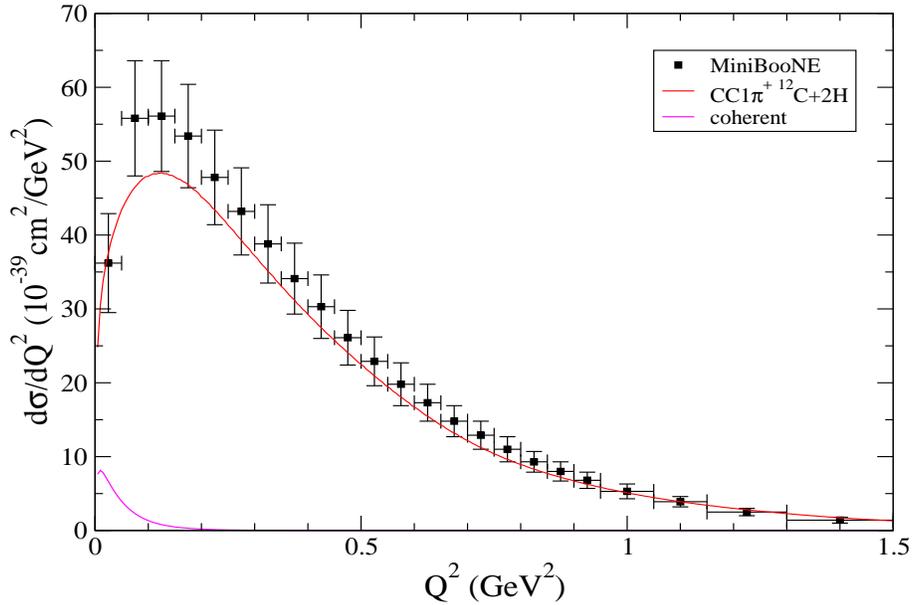}
\caption{(color online). MiniBooNE flux-averaged CC 1$\pi^+$ $\nu_\mu$-CH$_2$ $Q^2$ distribution. The coherent channel is separately shown.
The experimental MiniBooNE points are taken from Ref. \cite{AguilarArevalo:2010bm}.}
\label{fig_minib_dQ2_piplus}
\end{center}
%\vspace*{+0.8cm}
\end{figure}

In previous works \cite{Martini:2009uj,Martini:2010ex,Martini:2011wp,Martini:2013sha} 
we have investigated the quasielastic channel measured by MiniBooNE \cite{AguilarArevalo:2010zc,AguilarArevalo:2013hm}. 
Here  we want to test our model, as described in Ref. \cite{Martini:2009uj}, 
also on the pion production channel. In this section we  compare the MiniBooNE data \cite{AguilarArevalo:2010bm} 
on  single charged pion production by neutrino reactions on mineral oil,
 CH$_2$, with our %flux integrated 
predictions based on our model of Ref. \cite{Martini:2009uj}. 
The results of the double differential cross section as a function of the muon variables, emission angle and energy, hence not affected by the neutrino energy reconstruction problem,  are shown in Fig. \ref{fig_minib_d2s_piplus}. 
Our theoretical cross section 
for the molecule CH$_2$ incorporates the two hydrogen contributions which are  
free of nuclear effects. The general agreement between our evaluation and the data is good. 
The single differential cross section as a function of the muon kinetic energy is
shown in Fig. \ref{fig_minib_ds_dt_piplus} and the $Q^2$ distribution in Fig. \ref{fig_minib_dQ2_piplus} in which the coherent contribution is also singled out. 
These quantities are also rather well reproduced.
Our model does not incorporate the final state interaction for the emitted pion 
on its way out the nucleus  which  reduces the pionic cross section and which should 
  lead  to an overestimation of our theory as compared to the data.
However this difference does not show up in this comparison with data. The role of final state interaction has been discussed by several authors  \cite{Lalakulich:2012cj,Hernandez:2013jka} without a clear experimental manifestation of its influence, a puzzle recently reviewed in Refs. \cite{Rodrigues:2014jfa,Alvarez-Ruso:2014bla}.

\begin{figure}
\begin{center}
  \includegraphics[width=12cm,height=8cm]{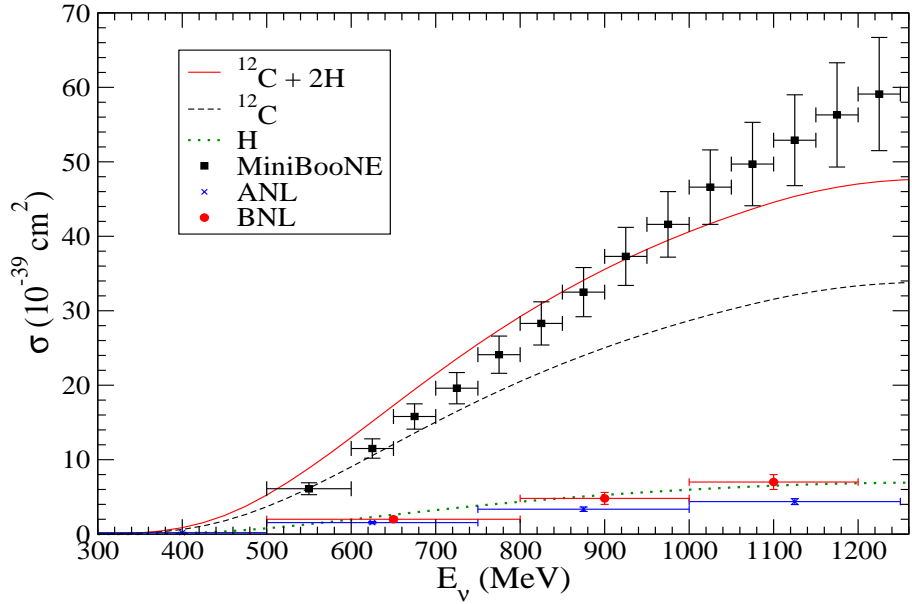}
\caption{(color online). CC 1$\pi^+$ $\nu_\mu$-CH$_2$ total cross section 
as a function of the neutrino energy compared to the experimental MiniBooNE results \cite{AguilarArevalo:2010bm}. Single-pion production on proton as well as ANL \cite{Radecky:1981fn} and BNL \cite{Kitagaki:1986ct} results are also given.}
\label{fig_x_tot_piplus}
\end{center}
%\vspace*{+0.8cm}
\end{figure}

The comparison of our total 
cross section with data for single charged pion production as a function of the neutrino energy  is  displayed in Fig. \ref{fig_x_tot_piplus}. 
In this figure our hydrogen contribution, which agrees with the BNL data \cite{Kitagaki:1986ct}, is  shown. The overall agreement is moderate, with a quality inferior to that of the double differential cross section. No correction has been applied in Fig. \ref{fig_x_tot_piplus} for the reconstruction of the neutrino energy which 
may increase further the  deviation of the theory from  the data \cite{Martini:2012fa,Martini:2012uc,Lalakulich:2012ac,Lalakulich:2012hs,Nieves:2012yz}. 
The  reason  for the inferior quality of the fit as compared to the double differential cross section is not clear.  At low energy the fact that the data are below the predictions could be an effect of the pion final state interaction which is not incorporated in our theoretical description. While the underevaluation above 1 GeV may be due to the two pion process. It does not enter directly in the single pion production process but it could influence it through a misidentification phenomenon, if one of the two pions is absorbed in the nucleus on its way out.  But in this case the reason for the success of the description of the double differential cross section remains unclear.  A similar question between the fits of the double differential cross section and the one of the integrated cross section  is present in the description of Ivanov \textit{et al.} \cite{Ivanov:2012fm}.

\begin{figure}
\begin{center}
  \includegraphics[width=12cm,height=8cm]{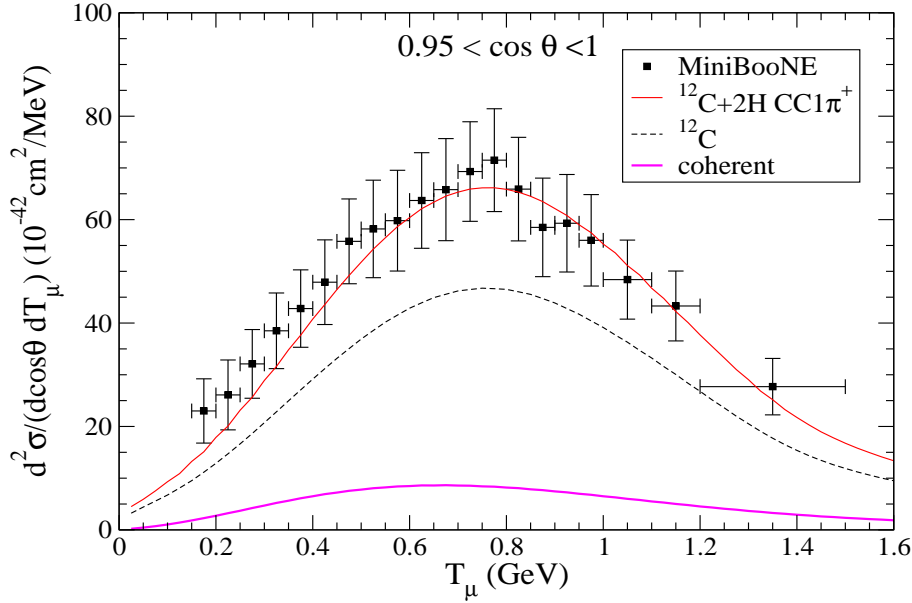}
\caption{(color online). MiniBooNE flux-averaged CC 1$\pi^+$ $\nu_\mu$-CH$_2$ double differential 
cross section for $0.95< \cos\theta <1$
as a function of the muon kinetic energy. The coherent channel is separately shown. 
The experimental MiniBooNE points are taken from Ref. \cite{AguilarArevalo:2010bm}.}

\label{fig_minib_d2s_piplus_cos0975m}
\end{center}
%\vspace*{+0.8cm}
\end{figure}  
It is interesting to display the results for the most forward bin for the muon angle $0.95< \cos\theta <1$. 
The double differential cross section for this bin is shown as a function of the muon kinetic energy in Fig. \ref{fig_minib_d2s_piplus_cos0975m}. 
Here the coherent contribution is significant although not dominant. 
This contribution is interesting due to its relation to a high energy collective state of the nucleus, denoted the pion branch \cite{Sawyer:1980ma, two_level}. 
It is a coherent mixture of Delta hole states and pions, as was nicely illustrated in the two level model of Delorme and Guichon \cite{two_level}. 
In the $(\omega, q)$ plane the pion branch, which embodies the modification of the dispersion relation for pion propagation 
in the nuclear medium by the polarization of the medium, i.e., by the virtual excitation of Delta-hole states, sits at lower energies than the free pion line. 
It  can only show up for probes which have, as the pion, a  spin longitudinal coupling, $\vec {S}\cdot \hat{q}$, i.e., along the momentum $\vec {q}$. 
In the interaction of physical pions with nuclei  it shows up only indirectly since the energy momentum relation, restricted to that of a physical pion, $\omega^2 -\vec{q}^2 = m_{\pi}^2$, is not that of the collective state. In this case one observes only the depletion due to the undetected collective state. 
The condition for its display in neutrino interaction was discussed by Delorme and Ericson \cite{Delorme:1985ps}.  
They pointed out that, for neutrinos, it is only in the forward direction that the spin longitudinal response which is sensitive to the pion branch, 
can dominate the cross section. This response contains the coherent pion production which represents the emission of a physical 
pion by a Delta-hole bubble, the nucleus remaining in its ground state. 
\begin{figure}
\begin{center}
  \includegraphics[width=12cm,height=1.8cm]{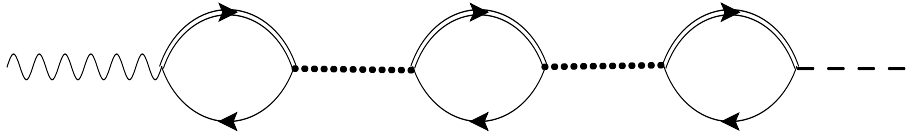}
\caption{Diagrammatic representation of the coherent pion production process. The wiggly line represents the external probe with the spin-longitudinal coupling. 
Double lines correspond to the propagation of a Delta, solid lines to the propagation of a nucleon hole, dotted lines to the Delta-hole interaction and 
the dashed line represents the pion.}
\label{fig_delta_coherent}
\end{center}
%\vspace*{+0.8cm}
\end{figure}  
 Pion emission can also occur via a series of Delta-hole bubbles, i.e., via the pionic decay of the pion branch in which the collective state transforms into a pion, as illustrated in Fig.\ref{fig_delta_coherent}. The signature for the collective pion branch is a shift towards smaller energy transfer of the coherent pion emission cross section with respect to the first order term with only one Delta-hole bubble. In the MiniBooNE experiment the  contribution of the coherent term is not sufficient to perform a quantitative study  but this possibility can be envisaged for the future. Notice that in analogy with the photoproduction of neutral and charged pion leading to discrete nuclear states \cite{Takaki:1985vm,Suzuki:1986nt}, also for the neutrino interactions a contribution from low energy excitations  of $^{12}$C or $^{12}$B in the case of charged currents, together with one pion emission, 
a ``quasi coherent'' pion emission, can also be expected in the forward bins.  It is not included in our description.

\section{Conclusion}
In summary we have tested our model of the neutrino nucleus interaction on the T2K inclusive data and on the MiniBooNE single pion production cross section. For the double differential cross sections which are free from neutrino energy reconstruction problems the agreement is generally satisfactory. 
The comparison with the T2K inclusive results represents the first successful test of the necessity of the multinucleon emission channel in an experiment 
with another neutrino flux with respect to the one of MiniBooNE. 
Even with the inclusion of the np-nh excitations, some underevaluation by the theory of the T2K data seems to show up  in the forward direction. 
It could be due to some contributions not included in our description, such as excitations of low-lying giant resonances. 
In the single pion production MiniBooNE data the lowest angle bin is sensitive to the coherent pion production cross section. Presently the importance of this contribution is not sufficient to perform a detailed study of this interesting channel but in the future it could become accessible with some improvements in the angular resolution so as to be more concentrated on the forward direction.
For the integrated cross sections  the underevaluation of our theory with respect to the data above an energy, $E_\nu \simeq 1$ GeV is presumably   due to the two pion production process which influences the inclusive cross section directly, but also the single pion exclusive one through a misidentification process if one of the two pions  is absorbed. The theoretical description should be improved in this direction with the inclusion of the two pion channel.
%\vskip 0.22 true cm
%\noindent{\bf{Acknowledgments}}
\begin{acknowledgements}
%\noindent 
M.M. thanks the TRIUMF laboratory for its hospitality and support during the ``Neutrinos and nuclear theory workshop'' and the Institute for Nuclear Theory at the University of Washington for its hospitality and the Department of Energy for support during the INT-13-54W Workshop.
We thank the organizers and the participants of these workshops for useful discussions. 
This work was partially supported by the Interuniversity Attraction Poles Programme initiated by the Belgian Science Policy Office (BriX network P7/12).
\end{acknowledgements}

\end{document}